\documentclass[aps,prl,twocolumn,groupedaddress]{revtex4}

\usepackage{graphicx}
\usepackage{bm}

\begin{document}

\title{Optical Activity of Planar Achiral Metamaterials}

\author{E. Plum}  \affiliation{Optoelectronics Research Centre, University of Southampton, SO17 1BJ, UK}

\author{V. A. Fedotov}
\affiliation{Optoelectronics Research Centre, University of
Southampton, SO17 1BJ, UK}

\author{N. I. Zheludev} \email[Email: ]{n.i.zheludev@soton.ac.uk}
\homepage[Homepage:]{www.nanophotonics.org.uk/niz/}
\affiliation{Optoelectronics Research Centre, University of
Southampton, SO17 1BJ, UK}

\date{\today}

\maketitle

\textbf{The phenomenon of optical activity that is the ability to
rotate the polarization state of light is a fundamental effect of
electrodynamics which is traditionally associated with mirror
asymmetry (chirality) of organic molecules, proteins and inorganic
structures. The effect has enormous importance for analytical
chemistry, crystallography, molecular biology, and the food industry
and is also a signature effect used to detect life forms in space
missions. The recognition of chirality as a source negative
refraction of light
\cite{Pendry1,Tretyakov2,Monzon,Jin,Cheng,Agranovich} needed for the
creation of a perfect lens \cite{Pendry2} inspired intense work in
developing microwave and optical artificial chiral metamaterials
\cite{Rogacheva,Wegener,Kuwata,Plum,Thiel}. In this paper we present
a somewhat surprising result that a very strong optical activity may
be seen in a metamaterial system consisting of meta-molecules that
itself are not chiral. Here chirality is drawn from the mutual
orientation of the wave propagation direction and two-dimensional
meta-molecule. We demonstrate the optical activity effect using an
artificially created non-chiral planar metamaterial structure and
show that it behaves indistinguishably from that seen for chiral
three-dimensional molecular systems manifesting resonant
birefringence and dichroism for circularly polarized electromagnetic
waves. Strong resonant optical activity is observed that is
accompanied by the appearance of a backward wave, which is
characteristic for negative-index media.}

\begin{figure}[h!]
\includegraphics[width=85mm]{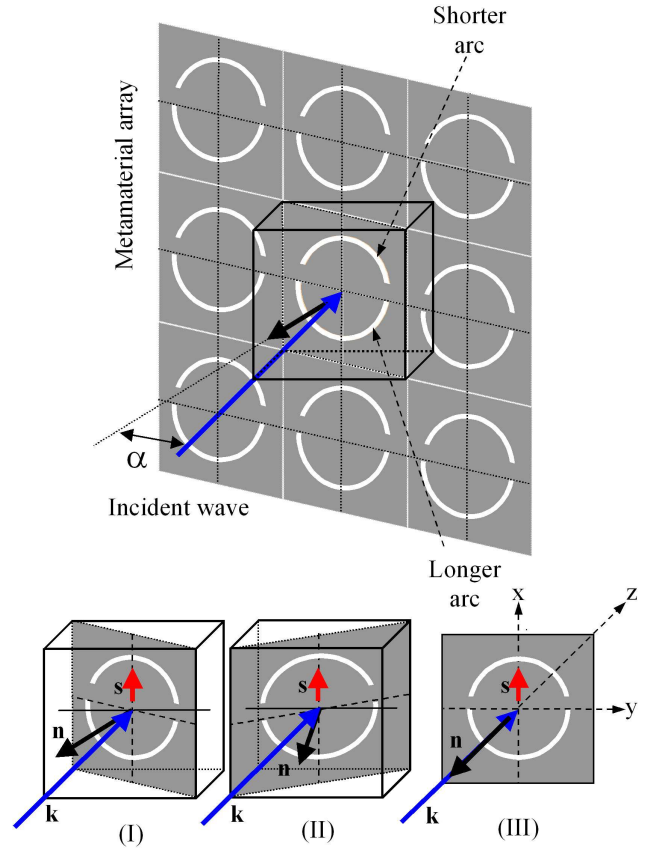}
\caption{\label{fig-structure} Planar slit metamaterial based on an
array of asymmetrically split rings that manifest optical activity
and circular dichroism at oblique incidence of light. The direction
of asymmetry is represented by a polar vector $s$ (directed from the
long to the short arc). Optical activity is seen when the
metamaterial plane is tilted around $x$-axis so the sample normal
$n$ is at an angle $\alpha \neq 0$ with the wave vector of the
incident wave $k$. Configuration I and II are the two enantiomeric
arrangements showing optical activity of opposite signs.
Configuration III corresponding to normal incidence shows no optical
activity.}
\end{figure}

\begin{figure}[h!]
\includegraphics[width=85mm]{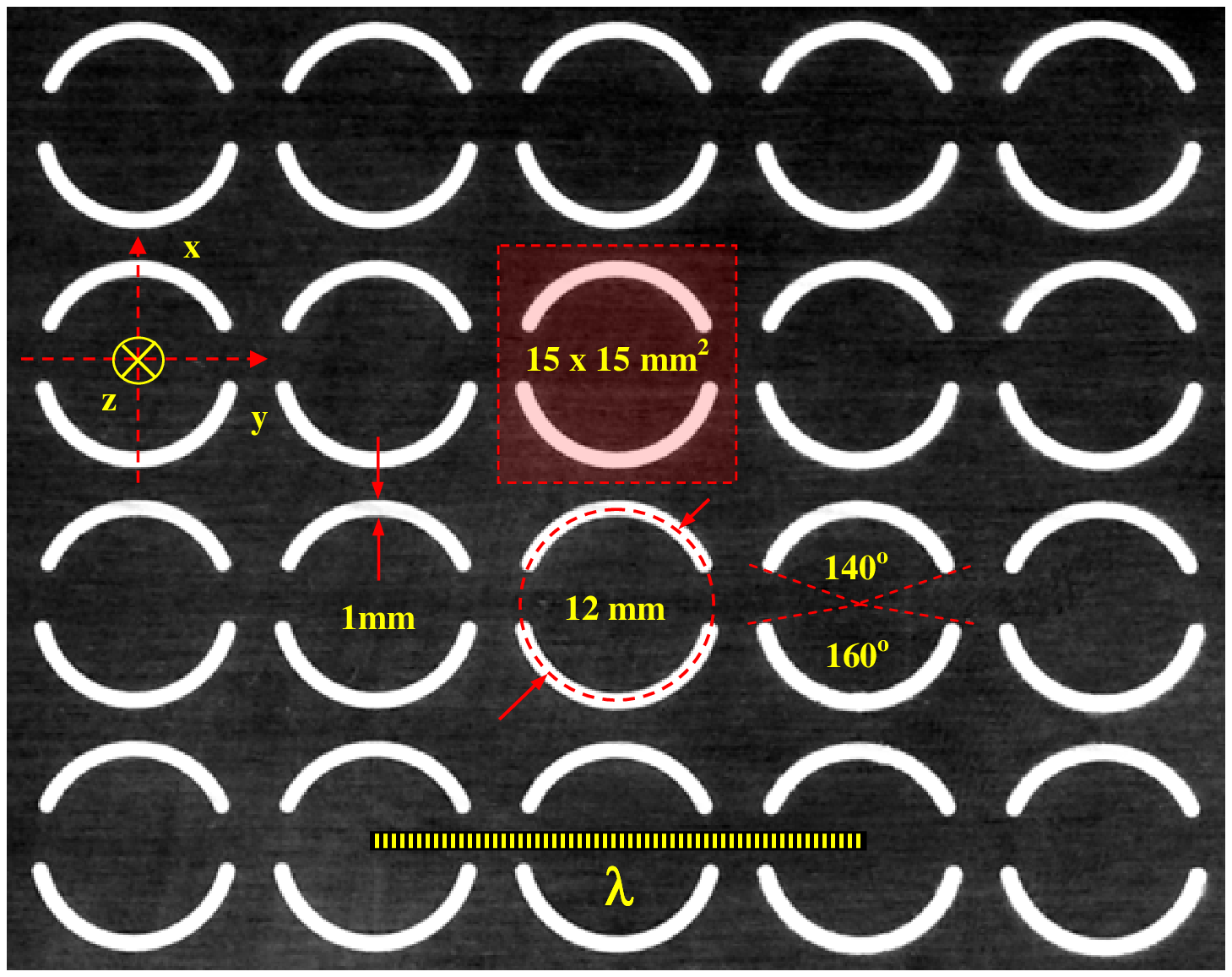}
\caption{\label{fig12} A fragment of the planar slit metamaterial
fabricated in $1~\text{mm}$ thick aluminium film photographed
against a light background. The highlighted square section shows the
elementary building block of the two-dimensional periodic
metamaterial structure. A bar in the lower part of the image
represents the wavelength $\lambda$ of radiation at which the
polarization resonance was observed.}
\end{figure}

Optical activity was first seen in 1811 by Dominique Arago and since
then, through the efforts of several generations of researchers was
linked to the three-dimensional property of molecules known as
chirality: a molecular structure such as a helix for which mirror
images are not congruent possesses a sufficient asymmetry to
manifest polarization rotation (optical activity). The effect of
optical activity is linked to the phenomenon of circular dichroism,
i.e. differential absorbtion for left and right circular
polarizations. The recent effort in creating artificial optically
active metamaterials that aimed to achieve strong optical activity
was focused on different types of arrays of 3D-chiral meta-molecules
\cite{Freeman,Engheta1,Engheta2,Lakhtakia1,Rogacheva,Wegener,Kuwata,Plum,Thiel}.
It is significantly less acknowledged that optical activity can also
be seen when oriented non-chiral molecules make a chiral triad with
the wave vector of light. This mechanism of optical activity was
first described by Bunn \cite{Bunn} and detected in liquid crystals
\cite{Williams}. Here we show that this is a highly significant
mechanism of optical activity in metamaterials that can be seen as
essentially planar structures that possess neither 2D chirality
\cite{Papakostas,Asymmetry} nor 3D chirality, and which are much
simpler to fabricate than metamaterials based on arrays of 3D-chiral
meta-molecules.

\begin{figure}[t!]
\includegraphics[width=85mm]{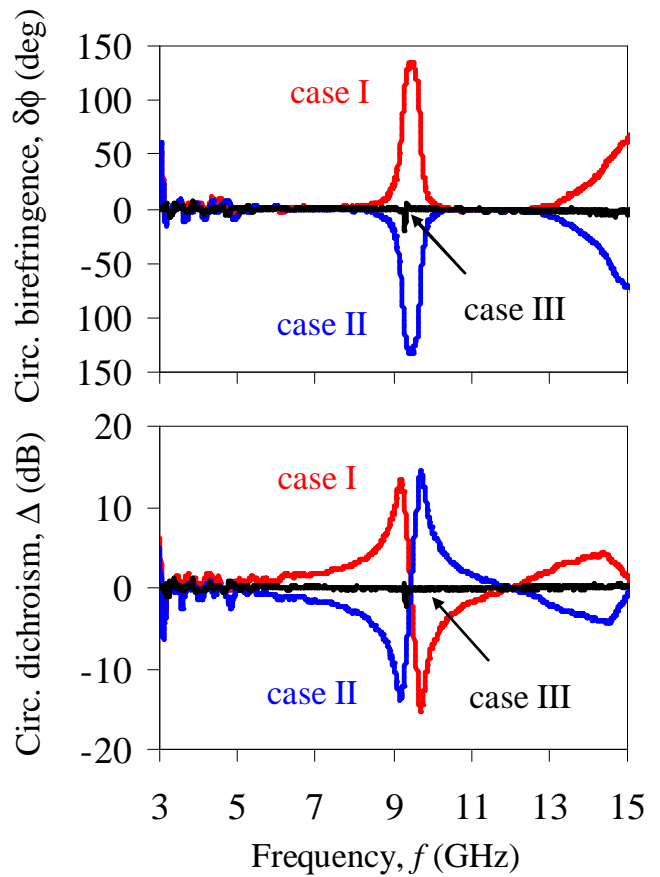}
\caption{\label{fig-gyrotropy} (a) Circular dichroism $\Delta$ and
(b) circular birefringence $\delta \phi$ of the planar metamaterial
structure measured at incidence conditions I, II (tilt angle $\alpha
= \pm 30^\circ$) and III ($\alpha = 0^\circ$) shown in Fig.~1.}
\end{figure}

We argue that to manifest optical activity meta-molecules of a
planar metamaterial structure may have a line of mirror symmetry,
but shall lack an inversion center, i.e. they shall posses a polar
direction $s$ (as illustrated in Fig.~1). A regular oriented array
of such meta-molecules will not show optical activity at normal
incidence. However, the metamaterial will become optically active at
oblique incidence provided that the plane of incidence does not
contain the polar direction. Indeed, in this case the wave vector
$k$, normal to the meta-molecule plane $n$ and polar vector $s$
constitute a three-dimensional chiral triad. The enantiomeric
configurations of these vectors corresponding to optical activity of
opposite signs are created by tilting the plane of the structure in
opposite directions with respect to the incident wave vector
(compare I and II in Fig.~1).

We observed optical activity in a self-standing thin metal plate
perforated with a regular two-dimensional array of ring slits
(Fig.~2). The ring slits are split asymmetrically into pairs of arcs
of different length separated by equal gaps. Each split ring has a
line of mirror symmetry along the $x$-axis but has no axis of
two-fold rotation, which enables us to introduce a polar vector $s$
that points in our case towards the short arc (see Fig.~1).

The origin of the effect  in such planar non-chiral structure may be
readily seen by considering a ``unit cell", which contains a section
of tilted metal plate with a single split ring slit (Fig.~1). Using
the terminology of crystallography, the direction of light
propagation will be a ``screw direction" of the unit cell (that is
to say it will have a sense of twist), if several conditions are met
\cite{Shaskolskaya}. First, the unit cell itself shall not have an
inversion center. This is assured by an asymmetry of the ring split.
Second, there should be no reflection symmetry in the plane
perpendicular to the propagation direction, which is provided by
oblique incidence. Third, there should be no inversion or mirror
rotation axis along the propagation direction. This is provided by
oblique incidence and asymmetric split. And finally there should be
no reflection symmetry for any plane containing the propagation
direction. This requirement is only fulfilled if the split is not
perpendicular, and therefore vector $s$ is not parallel, to the
incidence plane $yz$. Therefore, with reference to Fig.~1, in cases
I and II the direction of light propagation is a screw direction and
supports optical activity. On the contrary, case III, i.e. normal
incidence, fails the second, third and forth conditions of the
``screw direction" test. For instance at normal incidence there is a
plane of reflection symmetry containing the propagation direction.

In our experiments we measured losses and phase delays  for
circularly polarized electromagnetic waves propagating through the
metamaterial  (see Fig.~2). The slit structure has a number of
intriguing and useful properties. Being essentially a perforated
sheet of metal, it is not transparent to electromagnetic radiation
apart from a narrow spectral range around the resonant frequency, at
which the wavelength is approximately twice the length of the slits.
Transmission at the resonance is ``extraordinarily" high and
substantially exceeds that given by the fraction of the area taken
by the slits. As Joule losses in metals at these frequencies are
negligible, the incident energy is split between reflected and
transmitted radiation, and at the resonance reflection is low.
Around the resonance frequency and up to one octave above it the
structure does not diffract electromagnetic radiation: it becomes
diffractive for wavelengths shorter than the pitch of the array. As
will be illustrated below the structure shows a strong bell-shaped
resonance of circular birefringence leading to a strong polarization
rotation, while circular dichroism is at zero in the resonance. This
very useful feature is in striking contrast with optical activity in
most molecular systems, where characteristically strong resonant
polarization rotation is accompanied by substantial circular
dichroism resulting in elliptical polarization. Moreover at the
optical activity resonance the system shows no linear birefringence
(anisotropy) and eigenstates are therefore two circular
polarizations with equally moderate losses, making such a structure
an ideal platform for observing a negative index of refraction for
circular polarizations \cite{Pendry1}.

\begin{figure}[t!]
\includegraphics[width=85mm]{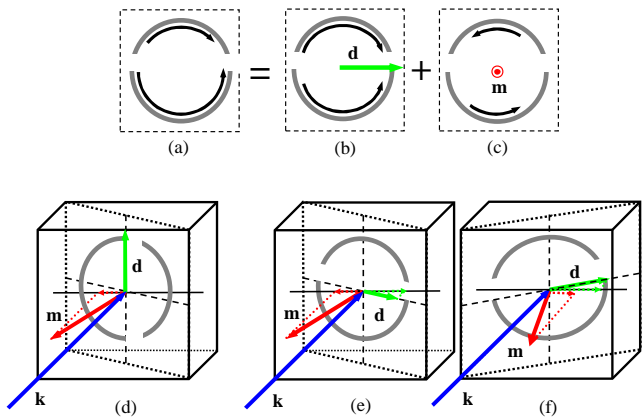}
\caption{\label{fig-orientation} Electric and magnetic responses in
an asymmetrically  split wire ring. Oscillating currents in the
split ring (a) can be represented as a sum of symmetric (b) and
anti-symmetric (c) currents that correspond to the induced electric
dipole in the plane of the ring $d$ (green arrow) and magnetic
dipole perpendicular to the plane $m$ (red arrow). For tilted
asymmetrically split rings polarization rotation is only absent if
the projections of $d$ and $m$ onto the plane perpendicular to the
$k$-vector (correspondingly green and red dashed arrows) are
orthogonal (d). If these projections are either parallel (e) or
anti-parallel (f), the strongest polarization rotation occurs, where
the optical activity for cases (e) and (f) has opposite signs.}
\end{figure}

If the structure is considered as a ``black box" the measurements of
losses and phase delays for circularly polarized electromagnetic
waves provide information on the circular dichroism and optical
activity of the medium in the ``black box". In practical terms we
measured the complex transmission matrix $t_{ij}$ for circularly
polarized waves. Here subscripts $+$ and $-$ denote left and right
circularly polarized waves correspondingly. Our measurements show
that the diagonal elements $(t_{++}$ and $t_{--})$ are generally not
equal indicating that the structure has true optical activity. The
difference between magnitudes of the diagonal elements $\Delta =
|t_{++}|^2 - |t_{--}|^2$ is a measure of circular dichroism of the
``black box", while the corresponding phase difference $\delta \phi
= \arg(t_{++}) - \arg(t_{--})$ is a measure of its circular
birefringence (see Fig.~3). The off-diagonal elements of the matrix
are equal within experiential accuracy, which indicates expected
presence of some anisotropy in the structure but also shows a
complete absence of the asymmetric transmission effect recently
discovered in planar chiral structures \cite{Asymmetry}. In all
cases experiments performed in opposite directions of wave
propagation show identical results.

\begin{figure}
\includegraphics[width=85mm]{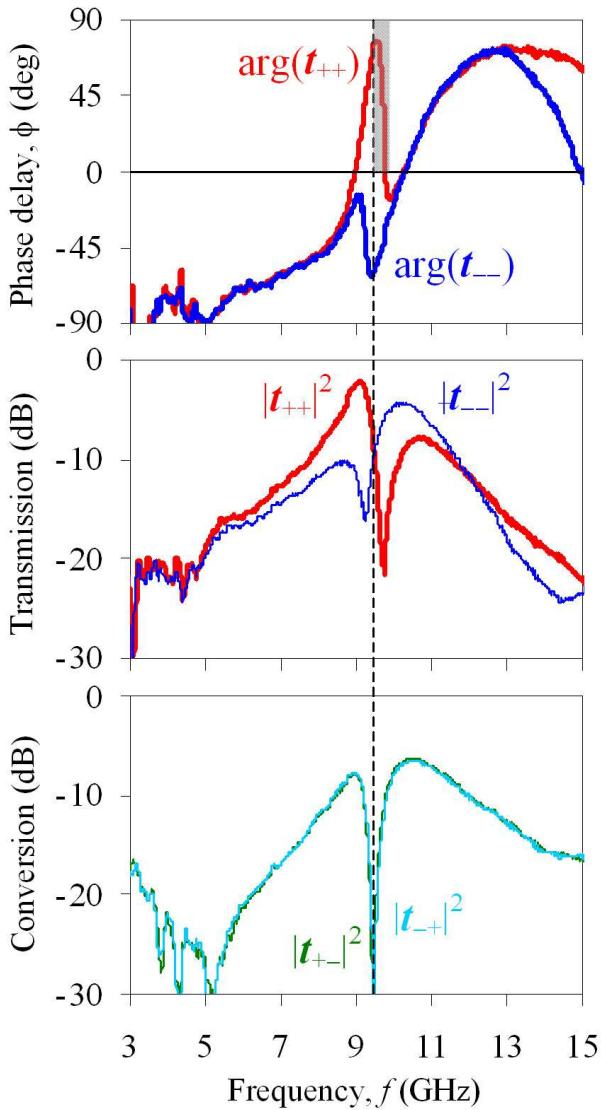}
\caption{\label{fig-5} (a) Dispersions of phase delay $\phi$ for the
transmitted left and right circularly polarized waves. The shaded
area indicates the frequency range with almost circular eigenstates,
where the phase velocity $v_p$ and group velocity $v_g$ for right
circular polarization have opposite signs, which is characteristic
for left-handed media. (b) Transmitted intensity of both left and
right circularly polarized waves. (c) The efficiency of circular
conversion, which is a direct indication of anisotropy (linear
birefringence) of the metamaterial response. The data in all panels
correspond to incidence condition I shown in Fig.~1, where the tilt
angle is $\alpha = 30^\circ$.}
\end{figure}

The following characteristic features of the effect have been
observed in the experiments: i) no circular birefringence or
dichroism is seen at normal incidence to the metamaterial array
($\alpha$ =0); ii) equal tilt in opposite directions yields circular
dichroism and circular birefringence of opposite sign. The observed
effect has a resonant nature. It is strongest around the resonance
between 9~GHz and 10~GHz, where the average arc length corresponds
to approximately half the wavelength.

The microscopic  origin of optical activity of the slit metamaterial
the can be easily understood by considering a complementary
structure, i.e. not the array of slits, but an array of metal wires
in the form of split rings (see Fig.~4). As with conventional
optical activity exhibited by chiral molecules, the effect must
result from the presence of both electric and magnetic responses.
Here, the structural asymmetry of the split rings plays a key role:
as illustrated in Fig.~4(a) a wave polarized along the split induces
unequal oscillating currents in the upper and lower arches of the
ring. This may be represented as a sum of symmetric and
antisymmetric currents that correspond to the induced electric
dipole in the plain of the ring and magnetic dipole perpendicular to
the ring (see Figs.~4(b) and (c)).


Now we shall consider non-normal incidence of the wave on the
structure (see Figs.~4(d)-4(f)).  Here blue, red and green arrows
represent the wave vector $k$ and induced magnetic $m$ and electric
$d$ dipoles of the metamaterial's unit cell, while dashed arrows
show projections of the corresponding dipole moments onto the plane
perpendicular to the wave vector. The magnetic dipole is always
oriented perpendicular to the plane of the structure and as we saw
above is only excited by the driving electric field along the split.
The structure shows optical activity if the split not perpendicular
to the plane of incidence. Maximum optical activity is observed when
the split is parallel to the plane of incidence, in this case the
wave vector and induced magnetic and electric dipoles are coplanar.
The mutual phase difference between the electric and magnetic
responses and thus the sign of optical activity depends on the sign
of the tilt (compare projections of electric and magnetic dipoles in
Fig.~4(e) and 4(f)). Similarly to how it happens in conventional
chiral media, when the wave vector and induced magnetic and electric
dipoles of the ``meta-molecule" are co-planar, dipole oscillations
create scattered electromagnetic waves with orthogonal polarizations
in the direction of wave propagation, and therefore the polarization
of the transmitted wave rotates. On the contrary, if the split is
perpendicular to the plane of incidence, the induced magnetic and
electric dipoles as well as their projections are orthogonal and the
structure does not show any optical activity (see Fig.~4(d)): the
oscillating magnetic and electric dipoles emit electromagnetic waves
of the same polarization that propagate along the direction of the
incident wave. According to Babinet's principle the slit
metamaterial structure that is complementary to the wire structure
discussed above, will exhibit similar polarization resonances in the
same frequency band.

In the slit metamaterial, at the resonance spectral band, from about
9 to $10~\text{GHz}$, anisotropy essentially disappears (here
circular conversion $t_{+-}=t_{-+} = \frac{1}{2}
\cdot(t_{xx}-t_{yy})$ is negligible, as shown in Fig.~5(c)). The
polarization eigenstates are very close to circular and in the
$k$-vector direction the material behaves as isotropic optically
active medium. Moreover, in this spectral range losses represented
by $|t_{++}|^2$ and $|t_{--}|^2$ are relatively small (see
Fig.~5(b)). Importantly, here phase velocity ($v_p \sim \omega /
\phi$, where $\omega=2\pi f$) and group velocity ($v_g \sim d\omega
/ d\phi$) for right circular polarization have opposite signs
indicating the appearance of a backward wave (see Fig.~5(a)).
The latter is evidence of left-handed material behavior and is a
signature of negative refraction in resonant bulk chiral media
\cite{Pendry1}. In accordance with Pendry's prediction negative
refraction should be seen at the resonance for one of the circular
polarizations only swapping to the opposite circular polarization
in the enantiomeric form of the medium. This is exactly what we
observed in our experiments: opposite signs of the group and phase
velocity are seen for right circular polarization at $\alpha =
30^\circ$ and for the left circularly polarized wave for the
enetiomeric arrangement, at $\alpha = -30^\circ$. Here the real
part of its effective refractive index can be estimated as $n
\approx - |c \phi / \omega h| \approx -2.5$, where $h$ is the
thickness of the structure in the direction of propagation.

In conclusion we have demonstrated strong resonant optical activity
using a non-chiral planar metamaterial, for which we also observed
signs of circularly polarized backward waves. We argue that stacking
such simple planar structures may provide an opportunity for
developing highly technologically relevant and practical negative
index media.

\begin{acknowledgments}
Financial support of the Engineering and Physical Sciences
Research Council, UK is acknowledged.
\end{acknowledgments}

\section{Methods summary}

\subsection{Sample description and manufacturing.} The planar
metamaterial is double-periodic with a square unit cell of
$15~\times~15~\text{mm}^2$ (see Fig. \ref{fig-structure}), which
ensures that the structure does not diffract electromagnetic
radiation at normal incidence for frequencies lower than 20~GHz. The
overall size of the sample was approximately
$220~\times~220~\text{mm}^2$. The metamaterial's unit cell contains
an asymmetrically split ring slit with the width of $1~\text{mm}$
and radius of $6~\text{mm}$ milled in a $1~\text{mm}$ thick
free-standing aluminum sheet.

\subsection{Measurement technique.} All the transmission measurements were performed in an anechoic chamber in $3 - 15~\text{GHz}$ range of frequencies using broadband horn
antennas (Schwarzbeck BBHA 9120D) equipped with lens concentrators
and a vector network analyzer (Agilent E8364B).





\begin{thebibliography}{21}
\expandafter\ifx\csname
natexlab\endcsname\relax\def\natexlab#1{#1}\fi
\expandafter\ifx\csname bibnamefont\endcsname\relax
  \def\bibnamefont#1{#1}\fi
\expandafter\ifx\csname bibfnamefont\endcsname\relax
  \def\bibfnamefont#1{#1}\fi
\expandafter\ifx\csname citenamefont\endcsname\relax
  \def\citenamefont#1{#1}\fi
\expandafter\ifx\csname url\endcsname\relax
  \def\url#1{\texttt{#1}}\fi
\expandafter\ifx\csname urlprefix\endcsname\relax\def\urlprefix{URL
}\fi \providecommand{\bibinfo}[2]{#2}
\providecommand{\eprint}[2][]{\url{#2}}

\bibitem[{\citenamefont{Pendry}(2004)}]{Pendry1}
\bibinfo{author}{\bibfnamefont{J.~B.} \bibnamefont{Pendry}},
  \bibinfo{journal}{Science} \textbf{\bibinfo{volume}{306}},
  \bibinfo{pages}{1353} (\bibinfo{year}{2004}).

\bibitem[{\citenamefont{Tretyakov et~al.}(2005)\citenamefont{Tretyakov,
  Sihvola, and Jylha}}]{Tretyakov2}
\bibinfo{author}{\bibfnamefont{S.}~\bibnamefont{Tretyakov}},
  \bibinfo{author}{\bibfnamefont{A.}~\bibnamefont{Sihvola}}, \bibnamefont{and}
  \bibinfo{author}{\bibfnamefont{L.}~\bibnamefont{Jylha}},
  \bibinfo{journal}{Photonics Nanostruct. Fundam. Appl.}
  \textbf{\bibinfo{volume}{3}}, \bibinfo{pages}{107} (\bibinfo{year}{2005}).

\bibitem[{\citenamefont{Monzon and Forester}(2005)}]{Monzon}
\bibinfo{author}{\bibfnamefont{C.}~\bibnamefont{Monzon}} \bibnamefont{and}
  \bibinfo{author}{\bibfnamefont{D.~W.} \bibnamefont{Forester}},
  \bibinfo{journal}{Phys. Rev. Lett.} \textbf{\bibinfo{volume}{95}},
  \bibinfo{pages}{123904} (\bibinfo{year}{2005}).

\bibitem[{\citenamefont{Jin and He}(2005)}]{Jin}
\bibinfo{author}{\bibfnamefont{Y.}~\bibnamefont{Jin}} \bibnamefont{and}
  \bibinfo{author}{\bibfnamefont{S.}~\bibnamefont{He}}, \bibinfo{journal}{Opt.
  Express} \textbf{\bibinfo{volume}{13}}, \bibinfo{pages}{4974}
  (\bibinfo{year}{2005}).

\bibitem[{\citenamefont{Cheng and Cui}(2006)}]{Cheng}
\bibinfo{author}{\bibfnamefont{Q.}~\bibnamefont{Cheng}} \bibnamefont{and}
  \bibinfo{author}{\bibfnamefont{T.~J.} \bibnamefont{Cui}},
  \bibinfo{journal}{Phys. Rev. B} \textbf{\bibinfo{volume}{73}},
  \bibinfo{pages}{113104} (\bibinfo{year}{2006}).

\bibitem[{\citenamefont{Agranovich et~al.}(2006)\citenamefont{Agranovich,
  Gartstein, and Zakhidov}}]{Agranovich}
\bibinfo{author}{\bibfnamefont{V.~M.} \bibnamefont{Agranovich}},
  \bibinfo{author}{\bibfnamefont{Y.~N.} \bibnamefont{Gartstein}},
  \bibnamefont{and} \bibinfo{author}{\bibfnamefont{A.~A.}
  \bibnamefont{Zakhidov}}, \bibinfo{journal}{Phys. Rev. B}
  \textbf{\bibinfo{volume}{73}}, \bibinfo{pages}{045114}
  (\bibinfo{year}{2006}).

\bibitem[{\citenamefont{Pendry}(2000)}]{Pendry2}
\bibinfo{author}{\bibfnamefont{J.~B.} \bibnamefont{Pendry}},
  \bibinfo{journal}{Phys. Rev. Lett.} \textbf{\bibinfo{volume}{85}},
  \bibinfo{pages}{3966} (\bibinfo{year}{2000}).

\bibitem[{\citenamefont{Rogacheva et~al.}(2006)\citenamefont{Rogacheva,
  Fedotov, Schwanecke, and Zheludev}}]{Rogacheva}
\bibinfo{author}{\bibfnamefont{A.~V.} \bibnamefont{Rogacheva}},
  \bibinfo{author}{\bibfnamefont{V.~A.} \bibnamefont{Fedotov}},
  \bibinfo{author}{\bibfnamefont{A.~S.} \bibnamefont{Schwanecke}},
  \bibnamefont{and} \bibinfo{author}{\bibfnamefont{N.~I.}
  \bibnamefont{Zheludev}}, \bibinfo{journal}{Phys. Rev. Lett.}
  \textbf{\bibinfo{volume}{97}}, \bibinfo{pages}{177401}
  (\bibinfo{year}{2006}).

\bibitem[{\citenamefont{Klein et~al.}(2007)\citenamefont{Klein, Wegener, and
  Linden}}]{Wegener}
\bibinfo{author}{\bibfnamefont{M.~D.~M.} \bibnamefont{Klein}},
  \bibinfo{author}{\bibfnamefont{M.}~\bibnamefont{Wegener}}, \bibnamefont{and}
  \bibinfo{author}{\bibfnamefont{S.}~\bibnamefont{Linden}},
  \bibinfo{journal}{Opt. Lett.} \textbf{\bibinfo{volume}{32}},
  \bibinfo{pages}{856 } (\bibinfo{year}{2007}).

\bibitem[{\citenamefont{Konishi et~al.}(2007)\citenamefont{Konishi, Sugimoto,
  Bai, Svirko, and Kuwata-Gonokami}}]{Kuwata}
\bibinfo{author}{\bibfnamefont{K.}~\bibnamefont{Konishi}},
  \bibinfo{author}{\bibfnamefont{T.}~\bibnamefont{Sugimoto}},
  \bibinfo{author}{\bibfnamefont{B.}~\bibnamefont{Bai}},
  \bibinfo{author}{\bibfnamefont{Y.}~\bibnamefont{Svirko}}, \bibnamefont{and}
  \bibinfo{author}{\bibfnamefont{M.}~\bibnamefont{Kuwata-Gonokami}},
  \bibinfo{journal}{Opt. Express} \textbf{\bibinfo{volume}{15}},
  \bibinfo{pages}{9575 } (\bibinfo{year}{2007}).

\bibitem[{\citenamefont{Plum et~al.}(2007)\citenamefont{Plum, Fedotov,
  Schwanecke, Zheludev, and Chen}}]{Plum}
\bibinfo{author}{\bibfnamefont{E.}~\bibnamefont{Plum}},
  \bibinfo{author}{\bibfnamefont{V.~A.} \bibnamefont{Fedotov}},
  \bibinfo{author}{\bibfnamefont{A.~S.} \bibnamefont{Schwanecke}},
  \bibinfo{author}{\bibfnamefont{N.~I.} \bibnamefont{Zheludev}},
  \bibnamefont{and} \bibinfo{author}{\bibfnamefont{Y.}~\bibnamefont{Chen}},
  \bibinfo{journal}{Appl. Phys. Lett.} \textbf{\bibinfo{volume}{90}},
  \bibinfo{pages}{223113} (\bibinfo{year}{2007}).

\bibitem[{\citenamefont{Thiel et~al.}(2007)\citenamefont{Thiel, Decker, Deubel,
  Wegener, and von Freymann}}]{Thiel}
\bibinfo{author}{\bibfnamefont{M.}~\bibnamefont{Thiel}},
  \bibinfo{author}{\bibfnamefont{M.}~\bibnamefont{Decker}},
  \bibinfo{author}{\bibfnamefont{M.}~\bibnamefont{Deubel}},
  \bibinfo{author}{\bibfnamefont{M.}~\bibnamefont{Wegener}}, \bibnamefont{and}
  \bibinfo{author}{\bibfnamefont{S.~L.~G.} \bibnamefont{von Freymann}},
  \bibinfo{journal}{Adv. Mat.} \textbf{\bibinfo{volume}{19}},
  \bibinfo{pages}{207} (\bibinfo{year}{2007}).

\bibitem[{\citenamefont{Tinoco and Freeman}(1957)}]{Freeman}
\bibinfo{author}{\bibfnamefont{I.}~\bibnamefont{Tinoco}} \bibnamefont{and}
  \bibinfo{author}{\bibfnamefont{M.~P.} \bibnamefont{Freeman}},
  \bibinfo{journal}{J. Phys. Chem.} \textbf{\bibinfo{volume}{61}},
  \bibinfo{pages}{1196} (\bibinfo{year}{1957}).

\bibitem[{\citenamefont{Jaggard and Engheta}(1989)}]{Engheta1}
\bibinfo{author}{\bibfnamefont{D.~L.} \bibnamefont{Jaggard}} \bibnamefont{and}
  \bibinfo{author}{\bibfnamefont{N.}~\bibnamefont{Engheta}},
  \bibinfo{journal}{Electron. Lett.} \textbf{\bibinfo{volume}{25}},
  \bibinfo{pages}{173} (\bibinfo{year}{1989}).

\bibitem[{\citenamefont{Mariotte and Engheta}(1995)}]{Engheta2}
\bibinfo{author}{\bibfnamefont{F.}~\bibnamefont{Mariotte}} \bibnamefont{and}
  \bibinfo{author}{\bibfnamefont{N.}~\bibnamefont{Engheta}},
  \bibinfo{journal}{Radio Sci.} \textbf{\bibinfo{volume}{35}},
  \bibinfo{pages}{827} (\bibinfo{year}{1995}).

\bibitem[{\citenamefont{Horn et~al.}(2004)\citenamefont{Horn, Pickett, Messier,
  and Lakhtakia}}]{Lakhtakia1}
\bibinfo{author}{\bibfnamefont{M.~W.} \bibnamefont{Horn}},
  \bibinfo{author}{\bibfnamefont{M.~D.} \bibnamefont{Pickett}},
  \bibinfo{author}{\bibfnamefont{R.}~\bibnamefont{Messier}}, \bibnamefont{and}
  \bibinfo{author}{\bibfnamefont{A.}~\bibnamefont{Lakhtakia}},
  \bibinfo{journal}{Nanotech.} \textbf{\bibinfo{volume}{15}},
  \bibinfo{pages}{303} (\bibinfo{year}{2004}).

\bibitem[{\citenamefont{Bunn}(1945)}]{Bunn}
\bibinfo{author}{\bibfnamefont{C.~W.} \bibnamefont{Bunn}},
  \emph{\bibinfo{title}{Chemical Crystallography}} (\bibinfo{publisher}{Oxford
  University Press}, \bibinfo{address}{New York}, \bibinfo{year}{1945}),
  p.~\bibinfo{pages}{88}.

\bibitem[{\citenamefont{Williams}(1968)}]{Williams}
\bibinfo{author}{\bibfnamefont{R.}~\bibnamefont{Williams}},
  \bibinfo{journal}{Phys. Rev. Lett.} \textbf{\bibinfo{volume}{21}},
  \bibinfo{pages}{342} (\bibinfo{year}{1968}).

\bibitem[{\citenamefont{Papakostas et~al.}(2003)\citenamefont{Papakostas,
  Potts, Bagnall, Prosvirnin, Coles, and Zheludev}}]{Papakostas}
\bibinfo{author}{\bibfnamefont{A.}~\bibnamefont{Papakostas}},
  \bibinfo{author}{\bibfnamefont{A.}~\bibnamefont{Potts}},
  \bibinfo{author}{\bibfnamefont{D.~M.} \bibnamefont{Bagnall}},
  \bibinfo{author}{\bibfnamefont{S.~I.} \bibnamefont{Prosvirnin}},
  \bibinfo{author}{\bibfnamefont{H.~J.} \bibnamefont{Coles}}, \bibnamefont{and}
  \bibinfo{author}{\bibfnamefont{N.~I.} \bibnamefont{Zheludev}},
  \bibinfo{journal}{Phys. Rev. Lett.} \textbf{\bibinfo{volume}{90}},
  \bibinfo{pages}{107404} (\bibinfo{year}{2003}).

\bibitem[{\citenamefont{Fedotov et~al.}(2006)\citenamefont{Fedotov, Mladyonov,
  Prosvirnin, Rogacheva, Chen, and Zheludev}}]{Asymmetry}
\bibinfo{author}{\bibfnamefont{V.~A.} \bibnamefont{Fedotov}},
  \bibinfo{author}{\bibfnamefont{P.~L.} \bibnamefont{Mladyonov}},
  \bibinfo{author}{\bibfnamefont{S.~L.} \bibnamefont{Prosvirnin}},
  \bibinfo{author}{\bibfnamefont{A.~V.} \bibnamefont{Rogacheva}},
  \bibinfo{author}{\bibfnamefont{Y.}~\bibnamefont{Chen}}, \bibnamefont{and}
  \bibinfo{author}{\bibfnamefont{N.~I.} \bibnamefont{Zheludev}},
  \bibinfo{journal}{Phys. Rev. Lett.} \textbf{\bibinfo{volume}{97}},
  \bibinfo{pages}{167401} (\bibinfo{year}{2006}).

\bibitem[{\citenamefont{Sirotin and Shaskolskaya}(1975)}]{Shaskolskaya}
\bibinfo{author}{\bibfnamefont{I.~I.} \bibnamefont{Sirotin}} \bibnamefont{and}
  \bibinfo{author}{\bibfnamefont{M.~P.} \bibnamefont{Shaskolskaya}},
  \emph{\bibinfo{title}{Crystal physics}} (\bibinfo{publisher}{Nauka},
  \bibinfo{address}{Moscow}, \bibinfo{year}{1975}), \bibinfo{note}{in Russian}.

\end{thebibliography}
\end{document}